\documentclass[prl,twocolumn,showpacs,superscriptaddress]{revtex4}%
\usepackage{graphicx}
\usepackage{dcolumn}
\usepackage{bm}

\begin{document}
\title{Metallic mean-field stripes, incommensurability and chemical potential in cuprates}
\author{J. Lorenzana}
\affiliation{Comisi\'{o}n Nacional de Energ{\'{\i }}a At\'{o}mica,
Centro At\'{o}mico Bariloche and Instituto Balseiro, 8400 S.C. de
Bariloche,
Argentina}
\author{G. Seibold}
\affiliation{Institut f\"ur Physik, BTU Cottbus, PBox 101344, 
         03013 Cottbus, Germany}
\date{\today}

\begin{abstract}
We perform a systematic slave-boson mean-field analysis of the three-band model
for cuprates with first-principle parameters. Contrary to widespread believe
based on earlier mean-field computations low doping stripes have a
linear density close to 1/2 added hole per lattice constant. We
find a dimensional crossover from 1D to 2D at doping $\sim 0.1$ 
followed by a breaking of
particle-hole symmetry around doping 1/8 as doping increases. Our
results explain in a simple
way the behavior of the chemical potential, the magnetic incommensurability,
and transport experiments as a function of doping. Bond centered and
site-centered stripes become degenerate for small overdoping.    
\end{abstract}
\maketitle
It is now a well established fact that doped holes in 
cuprates tend to self-organize in antiferromagnetic (AF) domain 
walls\cite{tra95,tra97,yam98,ara99,ara00}. 
 These quasi one-dimensional (1D) structures 
called stripes where predicted by mean-field theories\cite{zaa89,pol89hsch90}
inspired by the problem of solitons in conducting
polymers\cite{su79}.

In the superconducting systems observed stripes are parallel to the Cu-O bond 
(hereafter called  vertical stripes)
and  for low doping have a linear filling fraction $\nu \approx 1/2$ of added
holes per lattice constant ($a\equiv1$).
Unless a one-dimensional (1D)
instability opens a gap at the Fermi level\cite{zaa96}, half-filled ($\nu =
1/2$) stripes are expected to be metallic. 
 On the contrary mean-field theories predicted $\nu=1$ insulating 
stripes which led to an early rejection of stripes\cite{swche91}. 

Here we perform a systematic analysis of slave-boson (SB)
 mean-field\cite{kot86} solutions for a three-band Hubbard model as appropriate for the cuprates. 
Contrary to the early mean-field analysis we find that for low doping $x$ in a material like 
La$_{2-x}$Sr$_{x}$CuO$_{4}$ (LSCO) 
the most favorable stripes have indeed $\nu \approx
1/2$ (and are vertical) reconciling the mean-field picture with
experiment. Small doping introduces
more stripes at fixed $\nu$ reducing the stripe distance $d$ and 
 explaining the observed behavior of the magnetic incommensurability  $\epsilon
\approx x$ with $\epsilon=1/(2d)$\cite{tra95,tra97,yam98,ara99,ara00}. 

\begin{figure}[tbp]
\includegraphics[width=7cm,clip=true]{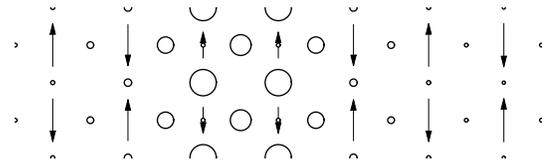}
\caption{BC vertical stripe with  $d=7$ and $\nu=1/2$ ($x=1/14$). 
Sites with (without) arrows are Cu (O). The length of the arrows is
 proportional to the local magnetization on Cu (negligibly on O).   
The radius of the 
circle represent the excess charge respect to the AF state (0.09 holes for
 the largest one). The $y$ axis is taken on the stripe direction
 (vertical). The picture repeats periodically in $x$ and $y$ directions}
\label{fig.rhos}
\end{figure}

We find low doping stripes to be centered on 
oxygen, bridging two Cu sites, hereafter named bond centered (BC)
stripes after the terminology of one-band 
models\cite{whi98,fle00} (see Fig.~\ref{fig.rhos}). 

\begin{figure}[tbp]
\includegraphics[width=7cm,clip=true]{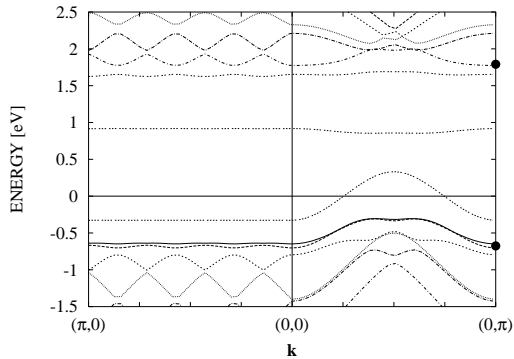}
\caption{Electron mean-field bands for the stripe of 
Fig.~\protect\ref{fig.rhos} measured from the chemical potential.
 Left (right) panel is in the direction perpendicular (parallel) 
to the stripe. 
 We also  plot the insulating charge transfer gap at momentum 
${\bf k}=(0,\pi)$ measured from the same reference energy (dots).} 
\label{fig.edk}
\end{figure}

Significant interstripe overlap sets in at $x\sim 0.1$ triggering a
crossover from 1D behavior at low doping to 2D behavior at large
doping. This does not affect the stability of stripes up 
to $x=1/8$ where a change of regime occurs.  Further doping occurs at
a fixed stripe distance $d=4$
and $\nu$ starts to grow. This explains the change of behavior
observed in both the incommensurability\cite{tra97,yam98,ara99,ara00}
 and the chemical potential\cite{ino97,har01} around doping $x=1/8$.
  Furthermore for $x<1/8$ the chemical potential is at the center of an
 approximately particle-hole symmetric 
band  (Fig.~\ref{fig.edk}) whereas for doping $x>1/8$ particle-hole 
symmetry is broken in good agreement with the picture deduced from transport 
experiments\cite{nod99,wan01}. Finally at $x\approx 0.21$ BC stripes 
and site-centered (SC) stripes become degenerate
 suggesting a regime of soft lateral stripe
fluctuations possibly relevant for superconductivity.  This 
quasidegeneracy was found in more sophisticated 
treatments of (less realistic)
one-band models\cite{whi98,fle00} however the ordering of the states was
reversed\cite{fle00}.

One-band models can provide qualitative understanding but may miss subtle
but important details specific for the cuprates. Therefore we use a
three-band Hubbard model with
a parameter set deduced from a first-principle LDA
computation for LSCO\cite{mcm90}. 
We view our computation as the last stage of a first principle
computation and thus without free parameters. 
The largest repulsion $U_{d}$\cite{mcm90} is treated within a 
SB  approach\cite{kot86} fully equivalent to the Gutzwiller 
approximation
and the other interactions are treated within the Hartree-Fock (HF)
approximation (an equivalent formalism has been used in Ref.~\cite{sad00} 
for electron-doped cuprates). 
We solve the mean-field equations both in real and in
momentum space. The real space computations are completely 
unrestricted\cite{goe98} except for the spins which are assumed 
to be  along the $z$ axis.

Typically we minimize the energy in large clusters with $N_s$ stripes 
($N_s=2,4$) of a maximum length $L=30$. 
Depending on boundary conditions, size and filling, 
solutions with and without charge order in the stripe direction are
obtained. In general due to the 1D character of stripes one expects
that a 1D instability will develop in large systems at low doping  
 rendering the system insulating even for $\nu<1$. 
Indeed if superconductivity is suppressed by pulsed magnetic 
fields the system is insulating below a critical doping and 
a crossover temperature $T^*$\cite{boe96}. 
In this work for simplicity we concentrate 
on the metallic state and  restrict to solutions without 
charge order in the stripe direction. 
Since $T^*$ is quite small compared with the typical electronic
scales we can set the temperature to zero in our computations.

Fig.~\ref{fig.rhos} depicts the typical charge
and spin distribution for a BC vertical stripe.
Added charge is accumulated in the antiphase boundary 
between the AF regions and mainly resides on oxygen sites.  

In Fig.~\ref{fig.edk} we show the band structure for the
electrons in an extended zone scheme.
Two bands appear well inside the charge transfer gap.
The upper one at $\sim 0.9$ eV is more Cu like and quite flat 
(even in the stripe
direction) whereas the lower one has substantial dispersion, 
is mainly O like, and crosses the Fermi level. Counting both ingap
bands the $\nu=1/2$ system corresponds to $3/4$ hole ($1/4$ electron) band filling. 
The gap between the bands is related to the magnetic character
of the stripe. (It is absent in SC solutions
with a non-magnetic core.) 
The electronic structure close to the chemical potential 
is well represented by a half-filled cosine like band, hereafter
referred to as the active band.

The dispersion of the active band perpendicular to the stripe 
is quite flat indicating quasi 1D behavior. Small oscillations appear 
in this direction for $d\leq 5$ which can be identified  as a  
crossover interstripe distance from 1D to 2D behavior and corresponds
to a crossover doping $x=0.1$ for $\nu=0.5$. 

The band structure in the stripe direction (right panel) is quite similar 
in all BC stripes solutions regardless of $d$ which instead determines the
periodicity in the perpendicular direction (left panel).
Comparison with photoemission data requires matrix elements
considerations and will be presented elsewhere. Here we notice that   
the Fermi surface crossing at vertical momentum $ k_y\approx \pi/4$ has 
been recently observed for $d=4$  stripes\cite{zho99} and the flat
portion of the active band in the direction perpendicular to the
stripe correlates well with the flat bands observed in many 
cuprates around  ${\bf k}\approx (\pi,0)$\cite{kin94}. 

 \begin{figure}[tbp]
 \includegraphics[width=7cm,clip=true]{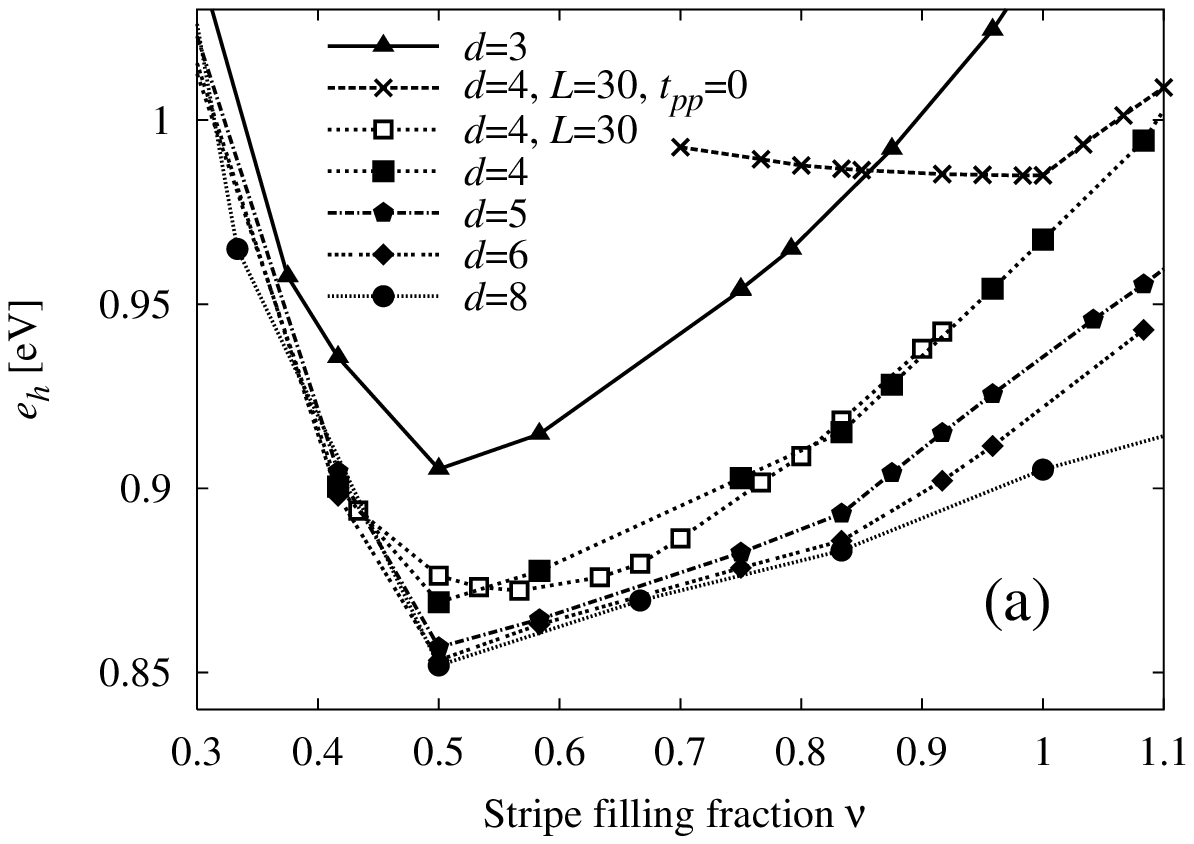}
 \includegraphics[width=7cm,clip=true]{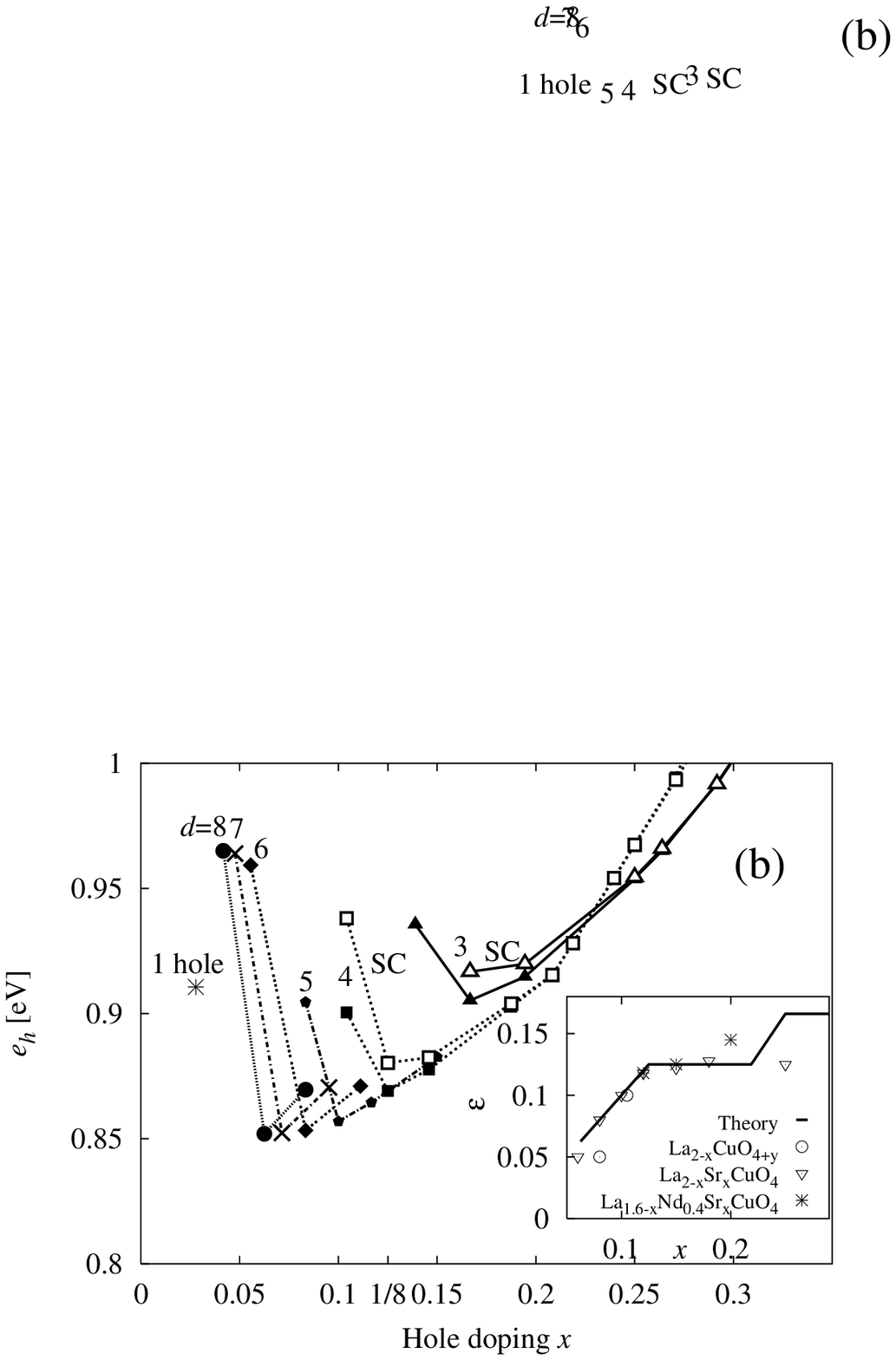}%
 \caption{$e_h$ as a function of $\nu$ (a) and $x$ (b) for vertical 
BC stripes. 
 Sizes are $4d\times 12$ for $d\leq 6$ and $2d\times 12$ for $d>6$.
 In (a) we also show the result for $4d\times 30$
 (open squares) and setting $t_{pp}=0$ shifted down by $0.89$
 eV (crosses). In (b) we also show the result
    for site-centered stripes  (labeled SC, open symbols) 
for $d=3,4$ and one self-trapped hole or ``electronic polaron'' 
at $x=1/36$. The inset in (b)
reports the incommensurability as obtained from the present calculation
(line) compared with experimental data from Ref.~\protect\cite{yam98}.}
 \label{fig.ehd}
 \end{figure}
 To evaluate the stability we compute the energy per added hole for 
$N_{h}$ holes added to the system with $N$ CuO$_{2}$ units,  
$e_{h}\equiv[E(N+N_{h})-E(N)]/{N_{h}}$.
 Here $E(N+N_{h})$ [$E(N)$]  is the total energy of the doped
  [undoped AF] solution.
 In Fig.~\ref{fig.ehd}~(a) we show $e_{h}$ as a function of the filling
 fraction $\nu \equiv N_{h}/(N_s L)$ for BC vertical stripes.
 Each curve corresponds to a fixed $d$. The curves have
 a sharp minimum at $\nu =0.5$. This is in contrast to early 
 one-band\cite{pol89hsch90,zaa96} and
 three-band mean-field computations\cite{zaa89} for which $\nu=1$ is 
 the most favorable hole filling. We can reproduce this behavior by 
 setting artificially $t_{pp}=0$ in our parameter set (crosses).
The difference in behavior can be understood by analyzing 
the band structure.
In the case $t_{pp}=0$ the active band becomes as narrow as the 
0.9eV band (Fig.~\ref{fig.edk}) and a gap develops between the 
active band and the band immediately bellow 
instead of overlapping as in Fig.~\ref{fig.edk}. This gap produces the
cusp singularity at $\nu=1$ [Fig.~\ref{fig.ehd}~(a)] 
via the discontinuity in ${\partial E}/{\partial N}$ 
which makes $\nu=1$ very stable. 
As $t_{pp}$ is increased the gap closes, the cusp 
singularity disappears, and a minimum develops in the $\nu<1$ 
part of the 
curve.  A simple model explains the formation of this minimum. 
Consider a 
flat density of states (DOS) of width $W$ to estimate the 
kinetic energy of the 
active band and neglect interactions.  In this approximation 
$e_h= e_0/\nu+W \nu/2$ where $e_0$ is  the energy cost per unit length to 
create a domain wall in the undoped system ($e_0\sim J/2$ with $J$ the 
superexchange constant). 
The optimum filling is given by  $v_{\rm opt}=\min(\sqrt{2 e_0/W},1)$. 
Increasing $t_{pp}$ increases 
$W$ without substantially affecting $e_0$  and produces a local minimum 
for  $\nu<1$. The stability close to $\nu=1/2$ is further enhanced
because  the DOS of a 1D  band is not flat but has a minimum at 
$\nu=1/2$.  For the same reason a gap produces a cusp in the energy, 
a minimum of the DOS enhances the local curvature of $e_h$ vs. 
$\nu$ and tends to shift the optimum filling closer to $\nu=1/2$.

We have  checked that for the most common parameter 
sets which contain $t_{pp}$\cite{mcm90}  
results are similar. However, as soon as $t_{pp}$ 
is set to zero, as in Ref.~\cite{zaa89}, half-filled stripes become 
unstable in favor of filled ones.



The $t_{pp}=0$ result is in accord with
HF computations in the simplest one-band model.
Including a second-neighbor hopping ($t'$) we have found that 
partially filled stripes {\em can be stabilized} also in one-band 
models by a similar mechanism, which will be reported elsewhere. 

Partially filled stripes where also found in the one-band Hubbard
model with $t'=0$ in a much stronger correlation regime and
going beyond a static mean-field\cite{fle00}. In our case, because
the active band is broad, correlations are less 
important and we expect that a static mean-field 
provides a good starting point. 



 Vertical stripes should be compared with other possible ground states.
 We find that within the present SB approach   
 they are lower in energy than polaron solutions\cite{lor93a}
 [see Fig.~\ref{fig.ehd} (b)] and diagonal stripes solutions (not
 shown).
Very low nonsuperconducting  dopings ($x<0.05$) were not explored in 
 detail since we believe that a careful consideration of other 
effects is required in this
 case.  Especially long-range order perpendicular to the planes
 favors loop configurations of stripes\cite{whi98}
 and long-range Coulomb effects\cite{low94,lor02} are
 expected to become important.

If one uses the  HF approximation instead of SB the minimum also 
occurs at $\nu\sim 1/2$ for BC vertical stripes. 
However a polaron lattice is the ground state in HF and 
diagonal stripes are  lower in energy than vertical ones. 
Thus the overall success of the present mean-field 
computation with respect to the earlier ones\cite{zaa89} 
is due to both a more  accurate mean-field approximation 
 SB instead of HF) and a more accurate parameter set.



 Further stabilization of the half-filled stripes 
 can occur if due to many-body effects
 a gap or pseudogap tends to open at the commensurate filling
 $\nu=1/2$\cite{zaa96}. 
 Although we do not need this effect to explain $\nu\approx
 1/2$ it is quite possible that this produces a fine tuning for
 $T<T^*$.  In this regard it is interesting to remark that
  the ``V'' shape form of the curves in Fig.~\ref{fig.ehd}~(a) 
 is due to the gap produced by the discretization of the levels in a
 finite system. For larger systems [see $d=4$ curve in
 Fig.~\ref{fig.ehd}~(a)] 
 the cusp becomes rounded, however the minimum is still 
 close to $\nu=1/2$. One should be aware that the curves with the
 cusp are for already quite long stripes ($L=12$) and it is not clear whether
 stripes in real materials will be much longer. Thus finite size
 data could turn out to be more realistic than infinite size one. 

 For $d>4$ all curves coincide close to $\nu=0.5$ in 
Fig.~\ref{fig.ehd}~(a)    
 whereas for $d=3,4$ the curves are shifted up. 
 This feature is due to the width of the 
 domain wall of $4\sim 5$ lattice sites (see
 Fig.~\ref{fig.rhos}) which forces stripes to overlap leading to an
 increase of $e_h$. 


 In Fig.~\ref{fig.ehd} (b) we show $e_{h}$ for various values of the 
 stripe separation $d$ and as a function of doping $x$ assuming each
 dopant introduces one hole. Since $x=\nu/d$ curves with larger 
$d$ appear at lower concentration. The locus of the minimum 
 of $e_{h}$ as a function of doping in Fig.~\ref{fig.ehd}(b) 
 is expected to form 
 a continuous curve in the thermodynamic limit by combining different 
 $d$ solutions in new solutions with larger periodicity. 
 Up to $x\approx 1/8$ the stripe filling is fixed at
 $\nu\approx 0.5$ and consequently the density of stripes increases
 with doping. This explains the behavior of the incommensurability
 $\epsilon=x/(2\nu)\approx x$ as seen in neutron scattering experiments
 in this doping range 
(inset of Fig.~\ref{fig.ehd})\cite{tra95,tra97,yam98,ara99,ara00}.
 For $x>1/8$ the right branch of the $d=4$ solution is more stable than
 the  $\nu\approx 0.5$ and $d=3$ solution due to the stripe overlap
 effect. Therefore the incommensurability remains locked  
 at $\epsilon \approx 1/8$ 
in good agreement with the change of behavior in
 $\epsilon$ observed around $x\approx 1/8$\cite{tra97,yam98,ara99,ara00}.

 As doping increases further, BC stripes become degenerate with 
SC ones at  $x\approx 0.21$. This suggests that
 as doping increases lateral fluctuations of the stripe will become
 soft and possibly mediate pairing between holes.


 For doping $x>0.225$ we find the $d=3$ stripe 
($\epsilon=1/6\approx 0.17$) 
 to become the lowest energy solution with an initial filling fraction 
 $\nu=0.675$.
 Experimentally the situation is not clear. $\epsilon=1/6$ has been 
 reported for
 YBa$_2$Cu$_3$O$_{6+\delta}$\cite{ara00} (YBCO) but not for 
 LSCO where $\epsilon$ remains in the 
 $\epsilon=1/8$ line up to $x=0.25$ (inset of Fig.~\ref{fig.ehd}). In 
La$_{2-x-y}$Nd$_{y}$Sr$_x$CuO$_4$ (LNSCO), where stripes are pinned
 by the low-temperature tetragonal lattice distortion,
 the incommensurability is substantially increased beyond $\epsilon=1/8$ 
 at doping $x=0.2$ but without reaching $\epsilon=1/6$\cite{tra97}.

 It is possible that lateral stripe fluctuations become 
 so strong at this doping range that an effectively isotropic state is 
 reached \cite{kiv98}.
 Another possibility is that the  $d=3$ phase is skipped 
 due to phase separation among the $d=4$ phase and the
 overdoped Fermi liquid which for our parameters becomes the lowest energy
 solution close to $x=0.4$ (see Ref.~\cite{uem01} for a related
 scenario). Further theoretical and experimental work should be
 done to clarify this point. Especially being the system charged  
 a careful analysis of phase separation is needed\cite{lor02}. 

 It has been recently emphasized\cite{har01} that there is a close
 connection between the doping dependent incommensurability as discussed
 above and the chemical potential in cuprates.
 In fact the chemical potential for the electrons can be related to
 $e_h$ via: 
 $\mu =-(e_{h}+x\frac{\partial e_{h}}{\partial x})$. From  
 Fig.~\ref{fig.ehd}(b) one can then deduce that $\mu$ is approximately
 constant for $x \lesssim 0.1$ and decreases for $x \gtrsim 0.1$ in 
 qualitative agreement with the observed behavior\cite{ino97,har01}.
 The rate of change of $\mu$ with doping, being a high derivative of
 the energy, is very sensitive to finite size effects and, moreover,  few
 experimental points are available in this doping range in order to allow for 
 a precise comparison. A rough estimate indicates that the theoretical 
 rate of change of $\mu$ with doping for $x>1/8$ is approximately a factor
 of 2 larger than the experimental one\cite{ino97,har01}. This may be due to
 phase separation among the $d=4$ stripe solution  and the paramagnetic 
overdoped Fermi liquid as mention above.

 We finally turn to the discussion of our calculation in light of 
 recent measurements of the Hall coefficient R$_{H}$ 
 in LNSCO\cite{nod99} and YBCO\cite{wan01}.
 In the former compound R$_{H}$
 displays an abrupt decrease below the charge-stripe ordering temperature
 T$_{0}$ and for concentrations $x \le 1/8$.   
 It has convincingly been argued that quasi
 1D transport is not enough to explain this anomaly and instead 
 reflects a remarkable cancellation due to particle-hole symmetry
 in the stripe state\cite{wan01,pre01}. 
 A partial suppression of R$_H$ (and simultaneously the 
 thermopower S) below some temperature T$_{max}$ 
 has also been observed in YBCO up
 to oxygen contents corresponding to doping $x\sim1/8$ \cite{wan01}.  
 For distant stripes we indeed observe that the chemical potential
 crosses an approximately particle-hole symmetric band 
 (see Fig.~\ref{fig.edk}). Interstripe hopping becomes
 significant for $x>0.1$ however it does not break particle-hole
 symmetry. 
 As discussed above the chemical potential shifts from the center of
 the band for $x>1/8$, thus breaking particle-hole symmetry and 
 consequently both $R_H$ and $S$ 
 start to grow in modulus. Unfortunately the magnitude and even 
 the sign of the transport coefficients  cannot be predicted by a
 knowledge of the band structure alone. For example  
 $R_H$ for our Fermi surface (which turns out to be open) 
 is determined by the unknown anisotropy of the scattering path 
 length\cite{ong91}.

 To conclude we have shown that, contrary to widespread believe,  
 an appropriate mean-field theory of stripe phases in cuprates is not
 in contradiction with
 experiments. Our theory explains in a simple way 
 the change of behavior of several experimental quantities
 (incommensurability, chemical potential, transport coefficients) around
 the ``magic'' doping $x=1/8$. Obviously the present stripes are 
 static whereas those in real materials are
 dynamic unless pinned by a structural distortion.
 Many properties
 however will be largely independent of the dynamical or static
 character of the stripes. The energy for example is determined by
 short range correlations and is insensitive to the long-distant behavior.  
 This is analogous to the problem of spin waves in a  1D
 antiferromagnet  where the spin wave approximation gives a reasonable 
 value for the energy although the system does not have true 
 long-range order\cite{and52}.
 Since the properties we have discussed are mainly determined by the
 energetics we believe that the present mean-field theory is a good
 starting point for an understanding of the electronic structure of
 hole doped cuprates.  


\end{document}